\documentclass[a4paper,twocolumn,11pt,accepted=2025-12-01]{quantumarticle}
\pdfoutput=1
\usepackage[utf8]{inputenc}
\usepackage[english]{babel}
\usepackage[T1]{fontenc}
\usepackage{amsmath}
\usepackage{hyperref}

\usepackage{lipsum}

\usepackage{amssymb}
\usepackage{soul}
\usepackage{dsfont}
\usepackage[utf8]{inputenc}
\usepackage{physics}
\usepackage{dsfont}
\usepackage{appendix}
\usepackage{xcolor}
\usepackage{graphicx,bbm}
\usepackage{mathtools}
\usepackage{cite}

\begin{document}

\title{Emergent Liouvillian exceptional points from exact principles}

\author{Shishir Khandelwal}
\affiliation{Physics Department and NanoLund, Lund University, Box 118, 22100 Lund, Sweden}
\orcid{0000-0003-3366-624X}
\email{shishir.khandelwal@teorfys.lu.se}
\author{Gianmichele Blasi}
\affiliation{Department of Applied Physics, University of Geneva, 1211 Geneva, Switzerland}
\orcid{0000-0002-4024-788X}
\email{gianmichele.blasi@unige.ch}

\maketitle

\begin{abstract}
 Recent years have seen a surge of interest in exceptional points in open quantum systems. The natural approach in this area has been the use of Markovian master equations. While the resulting Liouvillian EPs have been seen in a variety of systems and have been associated to numerous exotic effects, it is an open question whether such degeneracies and their peculiarities can persist beyond the validity of master equations. In this work, taking the example of a dissipative double-quantum-dot system, we show that exact Heisenberg equations governing system and bath dynamics exhibit the same EPs as the corresponding master equations. To highlight the importance of this finding, we prove that the paradigmatic property associated to EPs - critical damping, persists well beyond the validity of master equations. Our results demonstrate that Liouvillian EPs can arise from underlying fundamental exact principles, rather than merely as a consequence of approximations involved in deriving master equations. 
\end{abstract}

\section{Introduction}
Exceptional points (EPs) have emerged as a crucial property of non-Hermitian systems. Such systems naturally arise in open classical settings, for example, in optics \cite{Miri2019} and electronics \cite{DeCarlo2022}, and their connection with the fundamental topic of PT-symmetry \cite{ElGanainy2018} has further fueled interest in the topic. The progress on the classical and semiclassical fronts has led to the investigation of EPs in open quantum systems. While there are many approaches in this direction \cite{Muller2008,Sergeev2023,Saha2023,Archak2024a,Archak2024b,Lin2024}, the most common one has been the use of master equations (MEs). Due to its linear structure, the Lindblad ME can naturally be written as a homogeneous matrix differential equation, with a non-Hermitian coefficient or \textit{Liouvillian}\footnote{Throughout this work, we use the term "Liouvillian" strictly only for the generator of Lindbladian evolution.} matrix, which generally shows EPs \cite{AmShallem2015,Mathisen2018,Hatano2019,Minganti2019,Perina2022}. Liouvillian EPs have been recently explored in the contexts of topological properties \cite{Kumar2021,Chen2022,Abbasi2022,Sun2023,Bu2023,Khandelwal2024,Sun2024}, dynamics towards steady states \cite{Khandelwal2021,Zhang2022,Zhou2023,Chen2024,Zhang2024,Chatterjee2024},  postselection of quantum jumps \cite{Minganti2019,Minganti2020,Chen2021,Minganti2022} and entanglement production \cite{Kumar2022,Li2023,Teixeira2023,Khandelwal2024}. Since master equations constitute a fundamentally approximate approach, these investigations are limited in their regime of validity, specifically to weakly-coupled Markovian dynamics \cite{Breuer2007,Wiseman}. It was recently found that  non-Markovian effects can, in some scenarios, lead to entirely different EPs \cite{Lin2024}. However, it is an open question, whether the phenomena associated to Liouvillian EPs could carry over to regimes far beyond the validity of master equations. In this work, we ask the question: are Liouvillian EPs, and their associated effects, merely artefacts of master equation approximations, or do they reflect fundamental properties of open quantum systems?

We adopt a framework for exact solutions of dissipative fermionic systems \cite{Blasi2024,Rodriguez2024,Rolandi2023,Blasi2025}. The approach involves directly solving the Heisenberg equations for system and bath operators without making Born and Markov approximations which are essential for the Lindblad ME. While the solutions thus obtained are valid in arbitrary coupling regimes, the framework has a well-defined weak-coupling limit which has been shown to correspond exactly to the ME approach, and therefore forms a natural platform to investigate EPs beyond the ME. Counterintuitively, in this framework, it is possible to write the system dynamics through a non-Hermitian evolution matrix, a property that is typically associated to situations where bath degrees of freedom are traced out. Considering a dissipative system of two quantum dots, we show that a second-order EP naturally arises in the involved evolution matrix. Importantly, we show that there is an exact correspondence between the EP obtained using Heisenberg equations and the one obtained using the master equation. Crucially, by solving for exact dynamics, we analytically show that the key dynamical effect, critical damping, persists at this EP in the HE approach. Finally, we provide key hints that the same correspondence may hold for dissipative chains of quantum dots. Our results provide the first evidence that Liouvillian EPs can emerge from underlying fundamental principles, with implications extending far beyond previously understood regimes.

\begin{figure}[!htb]
\centering
\includegraphics[width=1\columnwidth]{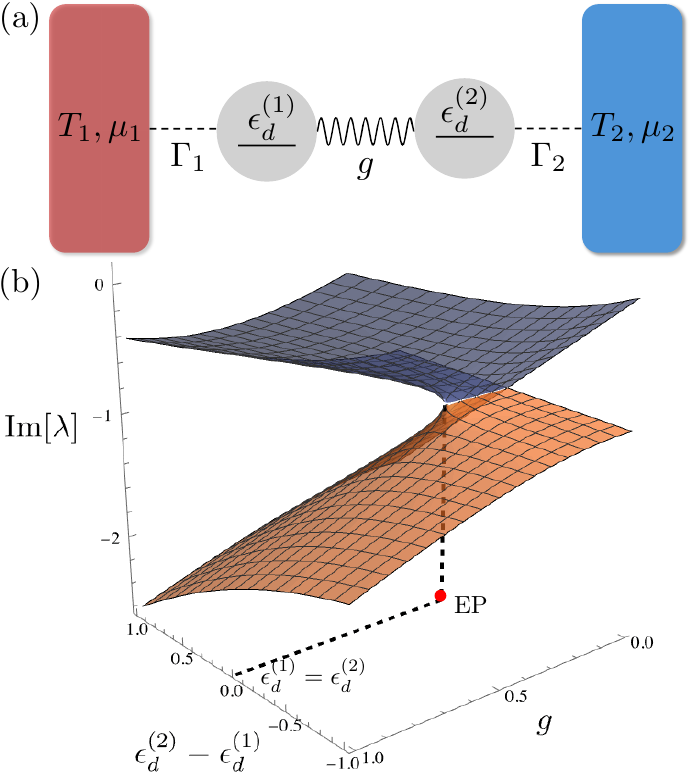}
\caption{(a) A two-terminal double quantum dot setup, with dot energies $\epsilon_d^{(j)}$, tunnel-coupling strength $g$ and reservoir couplings $\Gamma_j$ ($j=1,2$). (b) Riemann sheets corresponding to the eigenvalues of the Heisenberg evolution matrix $A$, in the space of the detuning ($\epsilon_d^{(2)}-\epsilon_d^{(1)}$) and $g$. The EP (depicted as a red dot) lies at zero detuning. We therefore consider resonant dots ($\epsilon_d^{(j)}\equiv\epsilon_d$) throughout this work.
}
\label{fig:setups}
\end{figure}

\section{Model} We consider a double quantum dot (DQD) setup, with each dot coupled to its own thermal reservoir of non-interacting fermions. The setup is depicted in Fig.~\ref{fig:setups} (a). The total Hamiltonian $\hat{H}$ is given by 
\begin{eqnarray}
\label{eq:totalH}
   \hat{H} =  \hat{H}^S + \sum_{j=1,2} \hat{H}^R_{j} + \sum_{j=1,2} \hat{H}^{SR}_{{j}}\,.
\end{eqnarray}$\hat H^S$ is the system Hamiltonian,
\begin{align}
\label{eq:system}
\hat{H}^S = \sum_{j=1,2} \epsilon_d \hat{d}_j^\dagger \hat{d}_j +g\left[\hat{d}_1^{\dagger}\hat{d}_{2}+\hat{d}_{2}^{\dagger}\hat{d}_1\right],
\end{align}where $\epsilon_d$ is the bare energy of the dots and $g$ is inter-dot coupling. The free fermionic Hamiltonian of reservoir $j$ is given by 
$\hat{H}^R_{j} = \sum_k \epsilon_{k j} \hat{c}_{k j}^\dagger \hat{c}_{k j}$, 
where $\hat c_{k j}^\dagger$ and $\hat c_{k j}$ are the creation and annihilation operators for the mode $k$ in reservoir $j$ ($j=1,2$). The dot and reservoir operators obey fermionic anti-commutation relations, $\{ d_i, d_{j}^\dagger\} = \delta_{ij}$ and $\{ c_{k  j}, c^\dagger_{k' j'}\} = \delta_{k k'} \delta_{j j'}$, respectively. Finally, the system-reservoir interaction Hamiltonian takes the form,
\begin{eqnarray}
\label{eq:H_SR}
   \hat{H}^{SR}_{{j}} =  \sum_{k } t_{k j}^{*}\hat{c}_{k j}^{\dagger}\hat{d}_j+t_{k j} \hat{d}^{\dagger}_j\hat{c}_{k j} \,,
\end{eqnarray}
where $t_{k j}$ represents the tunneling amplitude between the $j$-th quantum dot and the $k$-th mode of the corresponding reservoir.

We will tackle the problem of finding EPs in the dissipative DQD setup under two distinct frameworks. The first will involve considering Heisenberg equations (HE) for the evolution  of the dot and bath operators \cite{Blasi2024}. This approach will allow us to present an intuitive way of finding EPs and their effects in open quantum systems.  We will compare this with the usual approach involving the Lindblad master equation (ME) \cite{Khandelwal2021,Bourgeois2025}.   

\subsection{Heisenberg equation framework}\label{sec:hei}
In the Heisenberg framework, the evolution of the operators $\hat d_j$ and $\hat c_{k j}$ is given by the Heisenberg equations of motion ($\hbar,k_B = 1$),
\begin{align}\label{eq:Heisenberg_c}
\frac{d}{dt} \hat{d}_j = i [ \hat H, \hat{d}_j]\quad \text{and} \quad  \frac{d}{dt} \hat{c}_{k j} = i [ \hat H, \hat{c}_{k j}] 
\end{align}
In the solution to Eq. \eqref{eq:Heisenberg_c}, the bare tunneling rate is a key quantity, $\Gamma_{j}(\epsilon) = 2\pi \sum_{k}  \abs{t_{k j}}^2  \delta\left(\epsilon - \epsilon_{k j}\right)$. We operate in the wide-band limit (WBL), where its bandwidth exceeds all other energy scales in the system, allowing us to treat the tunneling rate as an energy-independent quantity, $\Gamma_{j}(\epsilon) \equiv \Gamma_{j}$ \cite{Brako1989,Baldea2016,Covito2018}. This is important to compare with the usual ME approach and to obtain closed-form solutions for the dynamics. It can be shown that the Heisenberg equations can be reduced to the following inhomogeneous equation (see the App. \ref{app:heis} for more details),
\begin{align}\label{eq:HEmatdiff}
\frac{d}{dt}\vec{\hat{d}}(t) = A \vec{\hat{d}} + \vec{\hat{ \xi}}
\end{align}where $\hat{\xi}_{j}(t) = -i\sum_{k}t_{k j}e^{-i\epsilon_{k j}(t-t_0)}\hat{c}_{k j}(t_0)\,$ and $\vec{\hat{d}} = (\hat d_1,\hat d_2)^T$. $A$ is a $2\times 2$ non-Hermitian matrix, that depends on system and reservoir parameters, taking the form, 
\begin{align}
    A = -\begin{pmatrix}
        \Gamma_1/2 +i\epsilon_d   &ig\\
        ig                       &\Gamma_2/2 + i\epsilon_d
    \end{pmatrix}.
\end{align}It is important to note that $A$ is mathematically neither a Liouvillian nor a non-Hermitian Hamiltonian, in the usual interpretation. It is not a Liouvillian as it is derived with no reference to ME approximations. Furthermore, it generates probability-preserving evolution with Eq. \eqref{eq:HEmatdiff}, unlike generic non-Hermitian Hamiltonians \cite{Minganti2019}. It has the following eigenvalues,
\begin{equation}
    \begin{aligned}
        \sigma(A)=\left\{-i\epsilon_d -\frac{\Gamma}{4} \pm \eta^{\text{\tiny{HE}}}\right\}
    \end{aligned}
\end{equation}and eigenvectors $(i\left(\Gamma_2-\Gamma_1 \pm \eta^{\text{\tiny{HE}}} \right)/4g,1)^T$, with $\eta^{\text{\tiny{HE}}} = \sqrt{\left(\frac{\Gamma_1-\Gamma_2}{4}\right)^2 -g^2}$. 
Clearly, at $\eta^{\text{\tiny{HE}}}=0$, the eigenvalues and eigenvectors merge. $\eta^{\text{\tiny{HE}}}=0$ is therefore, a second-order EP. We have chosen to consider only resonant dots, i.e., with the same energy $\epsilon_d$. It can be verified that this resonance is essential for the EP. We illustrate this in Fig. \ref{fig:setups} (b), taking off-resonant qubits, $\epsilon_d^{(1)}\neq\epsilon_d^{(2)}$. The Riemann sheets corresponding to the eigenvalues are shown, in the space of the detuning $(\epsilon_d^{(2)}-\epsilon_d^{(1)})$ and $g$.  As the plot shows, the EP is reached only at zero detuning.

\subsection{Master equation framework}
Under weak-coupling and Markov approximations the evolution of the dots can be described by a Lindblad master equation. Further, in the limit $g\ll\epsilon_d$ and $g\lesssim\Gamma_j$, dissipation can be described locally \cite{Hofer2017,Potts2021,Blasi2024} by an equation of the form, $\dot\rho(t) = \mathcal L\rho(t) $, with
\begin{equation}
\begin{aligned}\label{eq:ME}
    \mathcal L\rho(t) = -i\left[\hat H,\rho \right] + \sum_{j=1,2}&\Gamma_j(1-f_j\left(\epsilon_d \right)) \mathcal D\left[\hat\sigma^{(j)}_-\right]\rho\\&+\Gamma_jf_j\left(\epsilon_d \right)\mathcal D\left[\hat\sigma^{(j)}_+\right]\rho,
\end{aligned}
\end{equation}with the Fermi factor $f_j(\epsilon_d) = 1/(e^{(\epsilon_d-\mu_j)/T_j}+1)$ of reservoir $j$, characterized by temperature $T_j$ and chemical potential $\mu_j$, evaluated at the energy of the dots. The dissipator is defined as $\mathcal D[A] \rho   \coloneqq A\rho A^\dagger-(A^\dagger A\rho+\rho A^\dagger A)/2$. We have described the system under a Jordan-Wigner transformation \cite{Schaller2014} with $\hat H=\epsilon_d\sum_j\hat\sigma^{(j)}_+\hat\sigma^{(j)}_- +g\left(\hat\sigma_+^{(1)}\hat\sigma_-^{(2)} +\hat\sigma_-^{(1)}\hat\sigma_+^{(2)} \right)$, where $\sigma_{\pm}^{j}$ are raising and lowering operators. Although for clarity, we have chosen to work in the Schr\"odinger picture, one may equivalent solve the problem in the Heisenberg picture (see App. \ref{app:mas}) leading to the same eigenvalue structure and EPs. The Liouvillian $\mathcal L$ (restricted to the dynamically relevant steady-state subspace) is known to have the following eigenvalues \cite{Khandelwal2021}, 
\begin{align}\label{eq:eigs}
   \sigma(\mathcal L)= \left\{0,-\Gamma ,-\frac{\Gamma}{2},-\frac{\Gamma}{2},-\frac{\Gamma}{2} \pm2\eta^{\text{\tiny{ME}}} \right\}
\end{align} where $\eta^{\text{\tiny{ME}}} = \sqrt{\left(\frac{\Gamma_1-\Gamma_2}{4}\right)^2-g^2}$.  
There is an EP at $\eta^{\text{\tiny{ME}}}=0$, where the last three eigenvalues and their corresponding eigenvectors merge.  Importantly, the square-root factor is identical in the eigenvalues of both $\mathcal L$ and $A$, i.e., $\eta^{\text{\tiny{ME}}}=\eta^{\text{\tiny{HE}}}$. Therefore, the Liouvillian EP obtained with the Lindblad ME and the EP obtained in the HE framework overlap. The difference, however, lies in the order of the EP, second for HE and third for ME. The reason for the difference in orders can be understood at the level of transient dynamics and will be made clear in the next section. However, since the two EPs are characterised by the same square-root structure, it is clear that they are identical. We will henceforth drop the superscripts and refer to the square-root factor simply as $\eta$. 
 \begin{figure*}
    \centering
    \includegraphics[width=1\textwidth]{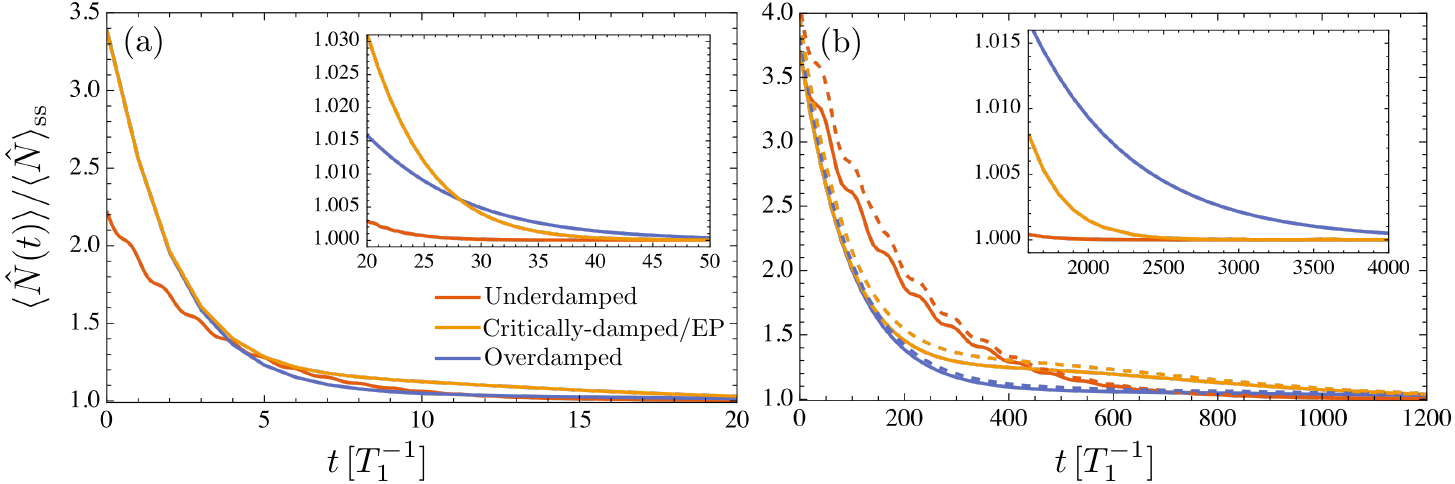}
    \caption{The population of dot 1 normalized by its steady state value, $\expval{\hat N_1(t)}/\expval{\hat N_1}_{\text{ss}}$, as function of time for (a) strong and (b) weak coupling, obtained with HE. The insets show the long-time behaviour. The dashed curves in (b) show master equation predictions. Common parameters: $T_1=1$, $T_2=0.1T_1$ $\epsilon_d= T_1$, $\mu_1=\mu_2=0$. Specific parameters: (a) $\Gamma_1 =0.5T_1$, $\Gamma_2=0.1T_1$, $g=3T_1$ (underdamping), $5\times 10^{-2}T_1$ (overdamping), $0.1T_1$ (EP) (b) $\Gamma_1 =10^{-2}T_1$, $\Gamma_2=10^{-3}T_1$, $g=5\times10^{-2}T_1$ (underdamping), $10^{-3}T_1$ (overdamping), $2.25\times10^{-3}T_1$ (EP).  }
    \label{fig:regimes}
\end{figure*}
\\



\section{EPs and dynamics in the two approaches} 
We have seen above that the EPs in the two approaches lie at the same point in parameter space. While this establishes connection between the spectra of the two approaches, the ME and HE approaches are fundamentally distinct and it is not obvious what the connection entails for the dynamics of the DQD.  
We sketch the dynamical solutions here and provide them in full detail in App. \ref{app:heis}. For simplicity, and without loss of generality, we focus on the populations of the dots, $\expval{\hat N_j(t)}\equiv \expval{\hat{d}^{\dagger}_j(t)\hat{d}_{j}(t)}$. It can be checked following a similar procedure presented here, that the same holds for all elements of the DQD density matrix individually, as well as for thermodynamic observables such as the current. We consider the evolution of the system and reservoir from time $t_0$ to $t$, with the initial occupations $n_j=\expval{\hat d_j^\dagger(t_0)\hat d_j(t_0)}$ with zero initial coherences and $f_j(\epsilon)=\expval{\hat c_j^\dagger(t_0)\hat c_j(t_0)}$. It can be shown (see also Ref. \cite{Blasi2024}) that the Heisenberg evolution \eqref{eq:HEmatdiff} can be solved for the transient population, leading to the following expression,
\begin{widetext}
\begin{equation}
\begin{aligned}
\label{eq:HEsol}
    \expval{\hat N_j(t)}=\sum_{m=1,2} D^*_{jm}(t)D_{jm}(t)n_m +\sum_{m=1,2}\Gamma_{m}\int \frac{d\epsilon}{2\pi} \tilde{D}_{jm}^*(\epsilon)\tilde{D}_{jm}(\epsilon)f_{m}(\epsilon),
\end{aligned}
\end{equation}\end{widetext}
with $D(t)\coloneqq e^{At}$ and
\begin{equation}
\label{eq:Dtilde}
\tilde{D}_{m m'}(\epsilon)=\int_{-\frac{t-t_0}{2}}^{\frac{t-t_0}{2}} ds D_{m m'}\left(\frac{t-t_0}{2}-s\right)e^{-i\epsilon s},
\end{equation}The above solution holds for both non-EPs and EPs. Eq. \eqref{eq:HEsol} consists of two parts. The first, initial-state-dependent part depends only on time $t$, i.e., it has a Markovian structure. It decays exponentially to zero in the steady state. The second part depends on the evolution at all times through the kernel $D\left((t-t_0)/2-s\right)$ in Eq. \eqref{eq:Dtilde}. It may remind the reader of non-Markovian master equations where memory effects are captured through a memory kernel. Indeed, this term can be shown to be the one that captures non-Markovian effects \cite{Blasi2024}. Furthermore, it is non-zero at all times, including the limit $t\to\infty$. Therefore, Eq. \eqref{eq:HEsol} contains non-Markovianity in both the transient and the steady state.  At non-EPs, $A$ is diagonalisable. As a result, its exponential can be written as a sum of purely exponential terms in time, $D(t) =\sum_i a_ie^{\lambda_it} \boldsymbol{v}_i$, where $\lambda_i$ and $v_i$ are eigenvalues and eigenvectors of $A$, respectively, and $a_i$ are scalars. However, at the $\eta=0$ EP, due to the non-diagonalisability of $A$, we have that $D^{\text{\tiny{EP}}}(t) = a_1e^{\lambda^{\text{\tiny{EP}}} t}\boldsymbol{v}^{\text{\tiny{EP}}} +a_2te^{\lambda^{\text{\tiny{EP}}} t}\boldsymbol{v}^\prime$, where $\lambda^{\text{\tiny{EP}}}$  and $\boldsymbol{v}^{\text{\tiny{EP}}}$ are the merged eigenvalue and eigenvector of $A$, respectively, and $\boldsymbol{v}^\prime$ is the generalised eigenvector \cite{Horn1985}. The appearance of a linear term in time along with a purely exponential one is characteristic of a second-order EP. Finally, due to the form of Eq. \eqref{eq:HEsol} with $D^*(t)D(t)$, the solution contains terms that come with $t^2$ along with a time-exponential factor.    

On the other hand, the solution to the ME \eqref{eq:ME} can be written as the exponential $\rho(t) = e^{\mathcal Lt}$. At non-EPs, this naturally translates to $\rho(t) = \sum_i c_ie^{\mu_i t}\hat\sigma_i$, where $\mu_i$ and $\hat\sigma_i$ are eigenvalues and eigenmatrices of $\mathcal L$, respectively. However, at $\eta=0$ there is a third-order EP, and we have $\rho^{\text{\tiny{EP}}}(t) = \sum_{i=1}^{3}c_i^{\text{\tiny{EP}}}e^{\mu_it}\hat\sigma^{\text{\tiny{EP}}}_i + (c_4^{\text{\tiny{EP}}}+c_5^{\text{\tiny{EP}}} t+c_6^{\text{\tiny{EP}}} t^2/2)e^{\mu^{\text{\tiny{EP}}}}\hat\sigma^{\text{\tiny{EP}}}+ (c_5^{\text{\tiny{EP}}} t+c_6^{\text{\tiny{EP}}}t)e^{\mu^{\text{\tiny{EP}}}}\hat\sigma^\prime+c_6^{\text{\tiny{EP}}}e^{\mu^{\text{\tiny{EP}}}}\hat\sigma^{\prime\prime}$, where $\hat\sigma^{\prime}$ and $\hat\sigma^{\prime\prime}$ are generalised right eigenmatrices of $\mathcal L$ \cite{Horn1985,Minganti2019,Khandelwal2021}. The $t^2$ factor arises due to a third-order EP. Therefore, we find that the HE and ME solutions both have $t^2$ terms, the former through a second-order EP and the latter through a third-order one. Through similar reasoning, it can be seen that a $n$-order EP in the HE should correspond to a $2n-1$-order EP in the corresponding ME. A similar observation has been made in the context of non-Hermitian Hamiltonians when considered in Hilbert space or in Liouville space \cite{Wiersig2020}. The difference here is the the HE and ME frameworks are distinct, and $A$ and $\mathcal L$ are not mathematically equivalent.

\section{Long-time dynamics, critical damping and the Mpemba effect}
We have seen above that the EPs obtained with the two approaches overlap and have the same structure. However, are their physical consequences the same? We show this for a paradigmatic EP effect on the dynamics.

As discussed above, the EP results in time-polynomial factors in the dynamics. While the effects of such terms can be observed at short times \cite{Cartarius2011,Khandelwal2021}, they also hold crucial importance at long times.  In Fig. \ref{fig:regimes}, we show the population dynamics for imaginary $\eta$ (underdamped, or oscillatory), $\eta>0$ (overdamped) and $\eta=0$ (EP) regimes, starting with the excited state of the two dots. The numerical calculation can be found at \cite{git}. In both weak (captured by both HE and ME) and strong coupling (captured only by HE), we see oscillations in underdamping, while smooth exponential decay in the other two regimes. Moreover, at long enough times, we find that the EP curves are closer to the steady state than the overdamped curves. This indicates that the EP is the point of critical damping, i.e., it represents the fastest non-oscillatory approach to the steady state. We now make this statement more precise. 

For the double quantum dot, it is known that the Liouvillian EP is the point of critical damping of the dynamics \cite{Khandelwal2021}. However, this result has been derived with a master-equation solution to the dynamics and its validity is limited to weakly-coupled Markovian systems. Here, we briefly sketch that a similar relation holds for exact dynamics of the double quantum dot, providing more details in App. \ref{app:B}.   
 We denote the average steady state population of dot $j$ by $\expval{\hat N_j}_{\text{ss}}$. Then, $\chi_j(t,\boldsymbol n) \coloneqq\left\lvert \expval{\hat N_j(t)}-\expval{\hat N_j}_{\text{ss}}\right\rvert$ is the absolute difference between the transient population from its steady-state value, with the initial populations given by the vector $\boldsymbol n=(n_1,n_2)$. We compare this distance at an EP (at $\eta=0$) and at a non-EP (at $\eta>0$, overdamping), i.e, we focus on the ratio $\mathcal R_j(t)=\chi_j^{\text{\tiny{EP}}}(t,\boldsymbol n^{\text{\tiny{EP}}})/\chi_j(t,\boldsymbol n)$, 
 where $\boldsymbol n^{\text{\tiny{EP}}}$ represents the initial populations in the case of critical damping, and $\boldsymbol n$ for overdamping. 
We note that different initial populations (i.e., $\boldsymbol n^{\text{\tiny{EP}}}\neq \boldsymbol n$) can be chosen for critical damping and overdamping within the ratio $\mathcal R_j(t)$, without affecting the following result. By extracting the exact solutions in the two regimes from Eq. \eqref{eq:HEsol} and then looking at the long-time behaviour, it can be shown that this ratio asymptotically approaches zero, behaving in the following manner,
\begin{align}
    \mathcal R_j(t)  \stackrel{\text{large t}}{\sim} \frac{\mathcal O(t^2)}{\mathcal O(e^{\eta t})}\stackrel{t\to\infty}{\rightarrow}0.
    \label{eq:Mpemba}
\end{align}
As a consequence, at long times, $\mathcal R_j<1$, which means that the state is closer to the steady state at the EP compared to any overdamped situation. We have obtained this result by varying only the inter-qubit coupling to interpolate between the overdamping and critical damping. The couplings  to the reservoir, which are the main determinant of the decay time, are kept the same for the two dynamical regimes. Notably, the above time-scaling is identical to the one found in \cite{Khandelwal2021} in the case of ME. 
Therefore, starting with arbitrary initial states, at long times, the relaxation to the steady state is faster at the EP than at in any overdamped situation, while the underdamped regime exhibits oscillations indefinitely. 
 Critical damping results in a phenomenon analogous to the counterintuitive quantum Mpemba effect \cite{Carollo2021,Chatterjee2023,Strachan2024,Chatterjee2024,Joshi2024,Rylands2023}: that quantum states that are initially further away from the steady state can relax faster towards it.

\begin{figure}
    \centering
    \includegraphics[width=1\columnwidth]{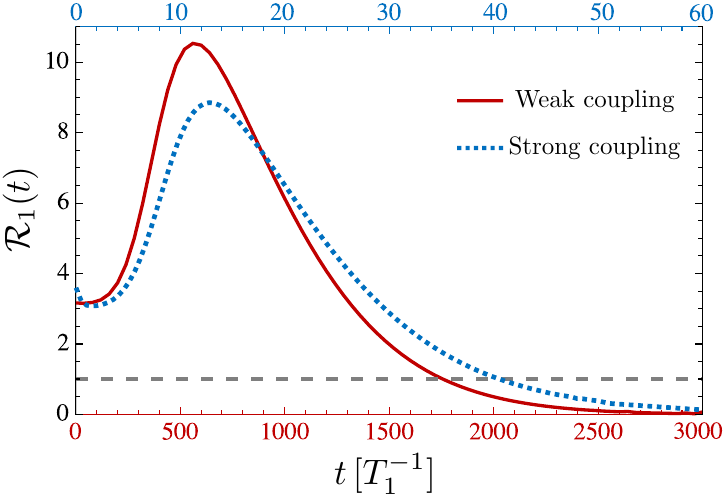}
    \caption{$\mathcal R_1$ as a function of time for strong and weak coupling, obtained with HE. $\mathcal R_1=1$ is marked with the dashed-gray line. The initial populations are chosen such that $\mathcal R_1>1$ at $t=0$. The parameters are taken from Fig. \ref{fig:regimes} (a) and (b), respectively. Similar plots can be obtained for $\mathcal{R}_2$.}
    \label{fig:Mpemba}
\end{figure}
Fig. \ref{fig:Mpemba} demonstrates this phenomenon, showing $\mathcal R_1$ as a function of time. A similar analysis would obviously work for $\mathcal R_2$. The initial states are chosen to be distinct for the two dynamical regimes - $\boldsymbol n^{\text{\tiny{EP}}}=(1,1)$ (i.e., the excited state for critical damping) and $\boldsymbol n=(0.5,0.5)$ (i.e., the maximally mixed state for overdamping). The same states are chosen for the two coupling regimes: strong (dashed curve) and weak (solid curve) in Fig. \ref{fig:Mpemba}. This ensures in our case that the system is further away from the steady state in the critically damped regime, i.e., $\mathcal R_1(0)>1$. At both weak and strong coupling, we find that at long enough times, $\mathcal R_1$ falls below 1, and goes exponentially to zero, as expressed by Eq. \eqref{eq:Mpemba}. We therefore find that critical damping is a faster approach to the steady state compared to overdamping, even if the system is initially further away from the steady state.

\section{Beyond the DQD model}Rigorously extending the above discussion to systems of more than two quantum dots, is in general a complex task. The simplest extension is a boundary-driven chain of three quantum dots, with equal inter-dot couplings $g$ and equal dissipation rates $\Gamma$ \footnote{For unequal dissipation rates, it can be checked that $A_{3}$ has cubic eigenvalues.} at the first and third dots. The Heisenberg evolution matrix $A_3$ has the eigenvalues 
\begin{align}\label{eq:3dot}
    \hspace*{-0.15cm}\sigma(A_{3}) = \left\{-i\epsilon_d-\frac{\Gamma}{2},-i\epsilon_d - \frac{\Gamma}{4}\pm\eta_3\right\}\hspace*{-0.1cm},
\end{align}with $\eta_3=\sqrt{2g^2-\left(\frac{\Gamma}{4}\right)^2}$, showing a second-order EP at $\eta_3=0$ or $g=\Gamma/4\sqrt{2}$. The corresponding local ME has among its eigenvalues $\{-5\Gamma/4\pm\eta_3,-3\Gamma/4\pm\eta_3\}$ (see App. \ref{app:C}). It therefore exhibits EPs at the same point ($\eta_3=0$) in parameter space as the HE. 
 
 The limiting factor to go beyond the above example is the lack of general closed-form expressions of eigenvalues. Specifically, for a chain of $N$ quantum dots with nearest-neighbour interaction, the Heisenberg evolution matrix $A_N$, is a $N\times N$ tridiagonal matrix, for which there are no such known closed-form expressions, in general. However, we note that this matrix also naturally exhibits EPs, and closed form expressions can be determined for specific cases; see App. \ref{app:C} for further details. On the other hand, calculating Liouvillian eigenvalues presents a similar hurdle \cite{Prosen2008,Prosen2010}. However, we expect that the consistency argument presented in this work demands a correspondence between EPs in the Heisenberg equations and suitably constructed master equations.

\section{Discussion} Through the analysis of exact solutions obtained from Heisenberg equations, we have shown that Liouvillian EPs can be seen as an inherent property of the model, not relying on the quantum master equation. Moreover, the EPs can result in similar effects on the dynamics; we demonstrated this with respect to critically damped dynamics towards the steady state, which results in a manifestation of the Mpemba effect. Crucially, our results point towards a fundamental nature of Liouvillian EPs, which extends the domain of their relevance in open quantum evolution. \par
We have focused on the "series" picture of the DQD model, with each dot connected to its own reservoir, in which we can sensibly define local dissipation. Interestingly, it can be checked that under global dissipation, i.e., when the dots interact with reservoirs globally \cite{Khandelwal2020,Khandelwal2021,Scali2021,Blasi2024}, neither the ME approach nor the HE approach show EPs, which further strengthens the connection between the two approaches. Beyond our idealised modeling of the dissipative DQD system, it would be interesting to consider the effect of environmental noise through (non-quadratic) dephasing-like interaction. In such a case, the Heisenberg framework can only be used perturbatively in the new coupling parameter. However, the treatment with master equations is straightforward \cite{Khandelwal2023c} and can give useful insights. In particular, for a small dephasing rate, we will still find an approximate critical damping situation with a modified scaling.
\par
Apart from critical damping that we have discussed in this work, it would also be of interest to see whether other physical effects associated with Liouvillian EPs also hold beyond the validity of the Lindblad equation, or whether there are some effects only arise in strongly coupled or non-Markovian regimes. For example, EPs are often associated to increased sensitivity to perturbations \cite{Wiersig2016,DeCarlo2022}. Since the EPs in both approaches have a characteristic square-root form, one may expect the same sensitivity (response to perturbations) but the response to quantum noise \cite{Wiersig2020,Wiersig2020a} will require careful analysis and may be different. Another direction of interest would be non-Hermitian state conversion in the vicinity of EPs. In the Lindbladian case, it is possible to see associated effects with a lowered fidelity than the non-Hermitian Hamiltonians \cite{Sun2023,Kumar2021,Chen2024,Khandelwal2024}. It would be of interest to see whether the fidelity can be increased within the HE framework. 
\par
While our work removes the Markovian and weak-coupling restrictions from the analysis of Liouvillian EPs, it keeps the wide-band limit. It is therefore an interesting open question to determine the precise conditions under which Liouvillian EPs can arise from exact principles, as well as to identify types of systems that can exhibit this property. Our work represents an initial step toward uncovering a general connection.

\section*{Acknowledgments}S.K. acknowledges support from the Swiss National Science Foundation grant P500PT\_222265 and the Knut and Alice Wallenberg Foundation through the Wallenberg Center for Quantum Technology (WACQT). G.B. acknowledges support from the Swiss National Science Foundation through the NCCR SwissMAP.



\onecolumn\newpage
\appendix

\section{Master equation for the double quantum dot: The Heisenberg picture}
\label{app:mas}

It is important to make a distinction between the Lindblad equation in the Heisenberg picture and the Heisenberg-equation approach (Sec. \ref{sec:hei} and App. \ref{app:heis}) used in this work. The former is a simple rephrasing of the usual Schrödinger-picture Lindblad equation, while the latter is a means of calculating the exact evolution of the dot and bath operators. 

As stated in the main text, one may equivalently solve the Lindblad equation for the DQD system without using the Jordan-Wigner transformation in the Heisenberg picture. In the Heisenberg picture the DQD (adjoint) Lindblad equation for the evolution of the operator $\hat A$ takes the form \cite{Breuer2007}
\begin{align}\label{eq:hei}
\dot{\hat{A}} = i\left[\hat H,\hat A \right] + \sum_{j=1,2}\Gamma_j(1-f_j\left(\epsilon_d \right)) \mathcal D^\dagger\left[\hat\sigma^{(j)}_-\right]\hat A +\Gamma_jf_j\left(\epsilon_d \right)\mathcal D^\dagger\left[\hat\sigma^{(j)}_+\right]\hat A,
\end{align}where the adjoint dissipators are defined as $\mathcal D^\dagger[\hat B]\hat A \coloneqq \hat{B}^\dagger \hat A\hat B - \{\hat{B}^\dagger \hat B,\hat A\}/2$. Solving the above equation requires the construction closed set of equations obtained for the expectation values of the dot operators. For fermions, such a construction is given by the so-called covariance matrix \cite{Prosen2008}, which contains the expectation values of products of Majorana modes of the system. For our DQD system, the situation is simplified by the form of the Hamiltonian and dissipators. We note that the dot populations can be expressed as $p_{11}=\expval{\hat n_1\hat n_2}$, $p_{10}=\expval{\hat n_1(\mathds 1-\hat n_2)}$, $p_{01}=\expval{(\mathds1-\hat n_1)\hat n_2}$ and $p_{00}=\expval{(\mathds 1-\hat n_1)(\mathds 1-\hat n_2)}$, while the coherences are corresponding to elements $\ketbra{10}{01}$ and $\ketbra{01}{10}$ are $\expval{\hat d_1^\dagger \hat d_2}$ and $\expval{\hat d_1\hat d_2^\dagger}$, respectively. Therefore, the most natural set of expectation values that can give a closed and complete set of equations is in the vector $\vec O =(\expval{\hat n_1},\expval{\hat n_2},\expval{\hat n_1\hat n_2},\expval{\mathds 1},\expval{\hat d_1^\dagger\hat d_2},\expval{\hat d_1\hat d_2^\dagger})$. Using Eq. \eqref{eq:hei}, we obtain the vectorised Lindblad equation, $\dot{\vec O} = L_{\text{H}} \vec O$, where 
\begin{align}
L_{\text{H}} = \left(
\begin{array}{cccccc}
 -\Gamma_1 & 0 & 0 & \gamma_1^+ & i g & -i g \\
 0 & -\Gamma_2 & 0 & \gamma_2^+ & -i g & i g \\
 \gamma_2^+ & \gamma_1^+ & -(\Gamma_1+\Gamma_2) & 0 & i g & -i g \\
 0 & 0 & 0 & 0 & 0 & 0 \\
 i g & -i g & 0 & 0 & -\frac{1}{2} (\Gamma_1+\Gamma_2) & 0 \\
 -i g & i g & 0 & 0 & 0 & -\frac{1}{2} (\Gamma_1+\Gamma_2) \\
\end{array}
\right). 
\end{align}The eigenvalues are given by 
\begin{align}
\sigma(L_{\text{H}}) = \{0,-\Gamma,-\frac{\Gamma}{2},-\frac{\Gamma}{2},-\frac{\Gamma}{2}\pm2\eta^{\text{\tiny{ME}}} \},
\end{align} where $\eta^{\text{\tiny{ME}}} = \sqrt{\left(\frac{\Gamma_1-\Gamma_2}{4}\right)^2-g^2}$. As expected, the eigenvalues of $L_{\text{H}}$ (i.e., the Liouvillian in the Heisenberg picture) are the same as the eigenvalues of the Liouvillian in the Schrödinger picture, presented in the main text (see Eq. \eqref{eq:eigs}). Furthermore, it can be checked that $L_{\text{H}}$ has a third-order EP at $\eta^{\text{\tiny{EP}}}=0$.

\section{Heisenberg equations of the double quantum dot}
\label{app:heis}
We utilise the framework developed in \cite{Blasi2024}. While we sketch the main aspects of the general framework here, further details can be found therein. The Heisenberg equations for the dot ($\hat{d}_j$) and reservoir ($\hat{c}_{kj}$) operators, with $j=1,2$, of the DQD system governed by the Hamiltonian in Eq. (1) in the main text are given by,
\begin{align}
	\label{diff_Eq_DQD_d_c}
	\frac{d}{d t} \hat{d}_j= i[\hat{ H}, \hat{d}_j]=-i\epsilon_d \hat{d}_j-i g \sum_{m\neq j} \hat{d}_m  -i \sum_{k } t_{k j}\hat{c}_{k j}, 
 \end{align}
 \begin{align}
	\label{diff_Eq_DQD_d_c_2}
	\frac{d}{d t} \hat{c}_{k j} = i[\hat{ H}, \hat{c}_{k j}]=-i \epsilon_{k j} \hat{c}_{k j}-i  t_{k j}^* \hat{d}_j\,.
\end{align}
Integrating Eq. \eqref{diff_Eq_DQD_d_c_2} and substituting into Eq. \eqref{diff_Eq_DQD_d_c}, 
\begin{equation}
	\begin{aligned}
		\hat{c}_{k j}(t)=e^{-i \epsilon_{k \alpha} (t-t_0)} \hat{c}_{kj}(t_0)-i \int_{t_0}^{t} ds~ e^{-i \epsilon_{k j}(t-s)} t_{k j}^* \hat{d}(s)
	\end{aligned}
\end{equation}and 
\begin{equation}
\begin{aligned}
	\frac{d}{d t} \hat{d}_j =&-i\epsilon_d \hat{d}_j -ig\sum_{m\neq j} \hat d_m-i \sum_{k} t_{k j} e^{-i \epsilon_{k j} (t-t_0)} \hat{c}_{kj}(t_0)\\&-\int_{t_0}^{t} ds~\sum_{k} \abs{t_{k j}}^2e^{-i \epsilon_{k j} (t-s)}\hat{d}_j(s).
\end{aligned}
\end{equation}Applying the wide-band limit $\Gamma_j\equiv\Gamma_{j}(\epsilon) = 2\pi \sum_{k}  \abs{t_{k j}}^2  \delta\left(\epsilon - \epsilon_{k j}\right)$, we obtain the following,
\begin{equation}
	\label{eq:diff_eq_d_component}
	\frac{d}{d t} \hat{d}_j=-\left(\frac{\Gamma_{j}}{2}+i\epsilon_d \right)\hat{d}_j-i\sum_{m\neq j} g\, \hat{d}_m + \hat{\xi}_{j}(t)\,,
\end{equation}
with $\hat{\xi}_{j}(t) = -i\sum_{k}t_{k j}e^{-i\epsilon_{k j}(t-t_0)}\hat{c}_{k j}(t_0)\,$. The above can be written as a matrix differential equation with the vectors $\vec{\hat{d}} = (\hat{d}_1, \hat{d}_2)^T$ and $\vec{\hat{\xi}} = (\hat{\xi}_1, \hat{\xi}_2)^T$,
\begin{equation}
	\label{eq:matrix_diff_eq}
	\frac{d}{d t}\vec{\hat{d}}(t)=A \vec{\hat{d}}(t)+\vec{\hat{\xi}}(t),  \quad A=-\begin{pmatrix}
		\frac{\Gamma_1}{2}+i \epsilon_d &  i g \\
		i g & \frac{\Gamma_2}{2}+i \epsilon_d 
	\end{pmatrix},
\end{equation}where $A$ is the non-Hermitian matrix that describes the evolution of the dots.

\subsection{Dynamics of the double quantum dot setup}
We focus on calculating the average occupation number or the population of the dots. In the wide-band limit, we can use the solution of Eq. \eqref{eq:diff_eq_d_component} to derive the following expression for the populations,
\begin{equation}
\begin{aligned}\label{eq:pop}
	\expval{\hat{d}^{\dagger}_j(t)\hat{d}_{j}(t)}=&\sum_{m=1,2} D_{jm}(t)^*D_{j m}(t)n_m\\&+\sum_{m=1,2}\Gamma_{m}\int \frac{d\epsilon}{2\pi} \tilde{D}_{jm}(\epsilon)^*\tilde{D}_{j m}(\epsilon)f_{m}(\epsilon),
\end{aligned}
\end{equation}with $D(t)\coloneqq e^{At}$ and
\begin{equation}
	\label{eq:Dtilde1}
	\tilde{D}_{j j'}(\epsilon)=\int_{-\frac{t-t_0}{2}}^{\frac{t-t_0}{2}} ds\, D_{j j'}\left(\frac{t-t_0}{2}-s\right)e^{-i\epsilon s}.
\end{equation}

\subsubsection{Dynamics at non-EP}
We consider the non-EP case with $\eta=\sqrt{\left((\Gamma_1-\Gamma_2)/4\right)^2-g^2}>0$. The corresponding transient solution has been previously considered in Ref. \cite{Blasi2024}. Here, we consider additional details, specifically ones relevant for our main results. When $\eta>0$, $A$ is diagonalisable  
\begin{align}
	D(t) = Se^{A_dt}S^{-1},\quad A_d = \begin{pmatrix}
		\lambda_1 &0\\
		0 &\lambda_2
	\end{pmatrix},
 \end{align}and
 \begin{align}
S= \left(
	\begin{array}{cc}
		-\frac{i (\Gamma_1-\Gamma_2+4\eta )}{4 g} & \frac{i (-\Gamma_1+\Gamma_2+4\eta )}{4 g} \\
		1 & 1 \\
	\end{array}
	\right)
\end{align}where $\lambda_{1,2}=-\frac{\Gamma}{4}\pm \eta-i\epsilon_d$.
\begin{equation}
	\begin{aligned} \expval{\hat{d}^{\dagger}_j(t)\hat{d}_{j}(t)} = \sum_{mpq=1,2}S^*_{jp}S^{-1*}_{pm}S_{jq}S^{-1}_{qm} \Bigg[& e^{\lambda^*_pt}e^{\lambda_q t} n_m \\&+ 4\Gamma_m e^{\left(\lambda^*_p+\lambda_q\right)\frac{t-t_0}{2}} \int \frac{d\epsilon}{2\pi}\frac{\sinh\left(\tilde\lambda_p^*\frac{t-t_0}{2} \right)}{\tilde\lambda^*_p}\frac{\sinh\left(\tilde\lambda_q \frac{t-t_0}{2} \right)}{\tilde\lambda_q}f_m(\epsilon)\Bigg],
	\end{aligned}
\end{equation}where we have defined $\tilde\lambda_p\coloneqq \lambda_p+i\epsilon$. In the steady-state, the term proportional to the populations naturally vanishes, while only one term in the integral survives. The final expression takes the form
\begin{equation}
\begin{aligned}
\expval{\hat{d}^{\dagger}_j\hat{d}_{j}}_{\text{ss}}&=\lim_{t\to\infty}\expval{\hat{d}^{\dagger}_j(t)\hat{d}_{j}(t)} \\&=\sum_{mpq=1,2}S^*_{jp}S^{-1*}_{pm}S_{jq}S^{-1}_{qm}4\Gamma_m \int \frac{d\epsilon}{2\pi} \frac{1}{\tilde\lambda_p^*\tilde\lambda_q}f_m(\epsilon).
\end{aligned}
\end{equation}As expected, the steady state is independent of the initial populations of the dots, $n_m$. However, it depends on the initial reservoir populations $f_m(\epsilon)$. 

\subsubsection{Dynamics at EP}
At the EP, $\eta=0$, or $g\equiv g_{\text{\tiny{EP}}}=\lvert\Gamma_1-\Gamma_2\rvert/4$. At this point, the eigenvalues of $A$ are $\lambda_1=\lambda_2=- \Gamma/4-i\epsilon_d \equiv\lambda$. The evolution matrix $D(t)$ is then given by $D^{\text{\tiny{EP}}}(t) =Te^{A_J t}T^{-1}$, with
\begin{align}\label{eq:mats}
 \quad A_J = \begin{pmatrix}
		\lambda &1\\
		0 &\lambda
	\end{pmatrix}\quad \text{and} \quad T = \left(
	\begin{array}{cc}
		-i & \frac{4 i}{\Gamma_1-\Gamma_2} \\
		1 & 0 \\
	\end{array}
	\right),
\end{align}where $A_J$ is the Jordan form of $A$ and $T$ is the corresponding transition matrix. Simple algebra leads to the following expression for the matrix elements of $ D^{\text{\tiny{EP}}}$,
\begin{align}\label{eq:DEP}
	D^{\text{\tiny{EP}}}_{jj^\prime}(t) = e^{\lambda t}\delta_{jj^\prime} +te^{\lambda t} T_{j,1}T^{-1}_{2,j^\prime}
\end{align}Using the above in Eq. \eqref{eq:Dtilde1}, we further find 
\begin{equation}\label{eq:DtEP}
	\begin{aligned}
		\tilde D^{\text{\tiny{EP}}}_{jj^\prime}(\epsilon) &= \int_{-\frac{t-t_0}{2}}^{\frac{t-t_0}{2}} ds~ \left[\delta_{jj^\prime} +\left(\frac{t-t_0}{2} -s\right) T_{j,1}T^{-1}_{2,j^\prime}\right]e^{\lambda (\frac{t-t_0}{2}-s)}e^{-i\epsilon s} \\
		& = 2e^{\lambda \frac{t-t_0}{2}}\left(  \mathcal F_1 \delta_{jj^\prime}  + T_{j,1}T^{-1}_{2,j^\prime} \mathcal F_2\right)
	\end{aligned}
\end{equation}where for convenience we have defined,
\begin{align}
	\mathcal F_1 = \frac{\sinh\big(\tilde \lambda \frac{t-t_0}{2}\big)}{\tilde \lambda}, \quad 
	\mathcal F_2 = \frac{t-t_0}{2}\frac{ e^{\tilde \lambda \frac{t-t_0}{2}}}{\tilde\lambda} - \frac{\sinh\big( \tilde \lambda \frac{t-t_0}{2}\big)}{\tilde\lambda^2},
\end{align}with $\tilde\lambda\coloneqq \lambda+i\epsilon$. Now, using Eqs. \eqref{eq:DEP} and \eqref{eq:DtEP} in Eq. \eqref{eq:pop}, we obtain after reshuffling terms,
\begin{equation}
	\begin{aligned}
\label{eq:popsum}&\expval{\hat{d}^{\dagger}_j(t)\hat{d}_{j}(t)}^{\text{\tiny{EP}}}= \left[1+t\left( T_{j,1}T^{-1}_{2,j} + T^*_{j,1}T^{-1*}_{2,j}\right)\right]n_je^{-\frac{\Gamma}{2}t} + \sum_m t^2 e^{-\frac{\Gamma}{2}t} n_m T^*_{j,1}T^{-1*}_{2,m}T_{m,1}T^{-1}_{2,j}\end{aligned}
\end{equation}\vspace*{-0.3cm}
\begin{equation*}
\begin{aligned}	&\quad\quad\quad+\int \frac{d\epsilon}{2\pi}4e^{-\frac{\Gamma}{2}\frac{t-t_0}{2}}\Big[ \Gamma_j f_j(\epsilon) \left\{\mathcal F_1^*\mathcal F_1 + T_{j,1}T^{-1}_{2,j} \mathcal F_1^*\mathcal F_2 + T^*_{j,1}T^{-1*}_{2,j} \mathcal F_2^*\mathcal F_1\right\} \\&\quad\quad\quad+ \sum_m \Gamma_m f_m(\epsilon) T^*_{j,1}T^{-1*}_{2,m}T_{m,1}T^{-1}_{2,j}\mathcal F_2^*\mathcal F_2
		\Big]
  \end{aligned}\end{equation*}The above can be simplified by using the exact expression for $T$ (Eq. \eqref{eq:mats}), specifically,
\begin{align}
	T_{j,1}T^{-1}_{2,j} = T^*_{j,1}T^{-1*}_{2,j} = g_{\text{\tiny{EP}}}\left( -1\right)^{\delta_{j,2}}, 
 \end{align}
 \begin{align}
 \sum_m T^*_{j,1}T^{-1*}_{2,m}T_{m,1}T^{-1}_{2,j}n_m =g_{\text{\tiny{EP}}}^2\left( n_1+n_2\right),
\end{align}where $g_{\text{\tiny{EP}}}=\lvert\Gamma_2-\Gamma_1\rvert/4$ is the inter-dot coupling at the EP. The population then simplifies to,
\begin{equation}
	\label{eq:popEPfinal}
	\begin{aligned}
	\expval{\hat{d}^{\dagger}_j(t)\hat{d}_{j}(t)}^{\text{\tiny{EP}}} &= \left(1+2t g_{\text{\tiny{EP}}} (-1)^{\delta_{j,2}}\right)n_je^{-\frac{\Gamma}{2}t} + t^2 g_{\text{\tiny{EP}}}^2(n_1 + n_2)e^{-\frac{\Gamma}{2}t} \\
		&+\int \frac{d\epsilon}{2\pi} 4e^{-\frac{\Gamma}{2}\frac{t-t_0}{2}}\Big[ \Gamma_j f_j\left(\epsilon \right) \left\{ \mathcal F_1^*\mathcal F_1 + g_{\text{\tiny{EP}}}(-1)^{\delta_{j,2}} \left (\mathcal F_1^*\mathcal F_2 + \mathcal F_2^*\mathcal F_1\right)\right\}    +g_{\text{\tiny{EP}}}^2\sum_m\Gamma_m f_m \mathcal F_2^*\mathcal F_2
		\Big]
	\end{aligned}
\end{equation}First, we note that there is a $t^2$ in time in the transient population Eq. \eqref{eq:popEPfinal}. This is due to the presence of a second-order EP. In general, for a $n$-th order EP in $A$, there will be a $t^{2(n-1)}$ term in the transient dynamics. To understand why this is the case, one may consider $A$ at a second-order EP, $A^{\text{\tiny{EP}}}$. The exponential $e^{A^{\text{\tiny{EP}}}t}$ naturally contains a linear factor in time along with exponential ones, due to the exponentiation of a Jordan form. In general, for an 
$n$-th order EP, $e^{A^{\text{\tiny{EP}}}t}$  contains factors of degree $n-1$, i.e., $t^{n-1}$. According to Eq. \eqref{eq:pop}, the population contains products of such exponentials, and therefore contains factors of $t^{2(n-1)}$.    Second, although it may seem that there are polynomial terms in time in the above expressions, for physical reasons, there cannot be any purely polynomial terms in time in the full transient solution, i.e., terms with a time-polynomial factor will necessarily exponentially decay to zero, as can be seen below in the long-time limit.

In the steady state $t\to\infty$, the population is given by
\begin{equation}
	\begin{aligned}
		\expval{\hat{d}^{\dagger}_j\hat{d}_{j}}^{\text{\tiny{EP}}}_{\text{ss}}&=\lim_{t\to\infty}\expval{\hat{d}^{\dagger}_j(t)\hat{d}_{j}(t)}^{\text{\tiny{EP}}} \\&= \int \frac{d\epsilon}{2\pi}\left[\Gamma_jf_j(\epsilon)\left\{\frac{1}{\left(\frac{\Gamma}{4}\right)^2 +\left(\epsilon-\epsilon_d \right)^2} +\frac{\Gamma g_{\text{\tiny{EP}}} (-1)^{\delta_{j,2}}}{2\left( \left(\frac{\Gamma}{4}\right)^2 +\left(\epsilon-\epsilon_d \right)^2\right)^2}\right\} 
		+ \frac{g_{\text{\tiny{EP}}}^2 \sum_{m}\Gamma_m f_m(\epsilon)}{\left( \left(\frac{\Gamma}{4}\right)^2 +\left(\epsilon-\epsilon_d \right)^2\right)^2}\right]
	\end{aligned}
\end{equation}

\section{Critical damping in the DQD}
\label{app:B}
We define the distance between the transient and steady-state populations,
\begin{align}
	\chi_j(t,\boldsymbol{n}) \coloneqq\left\lvert \expval{\hat{d}^{\dagger}_j(t)\hat{d}_{j}(t)}-\expval{\hat{d}^{\dagger}_j\hat{d}_{j}}_{\text{ss}}\right\rvert,
\end{align}where $\boldsymbol{n}=(n_1,n_2)$ is the vector of initial populations. Without loss of generality, we present the following result for $\chi_1$. The corresponding result for $\chi_2$ can be obtained following the same procedure. Keeping only the slowest decaying terms (i.e., ones decaying as $e^{-\Gamma t/2}$, while neglecting the ones decaying as $e^{-\Gamma t}$) in the above, we find the following long-time expression for $\chi_1$, respectively for the overdamped and the critical damped regimes,
\begin{equation}
	\begin{aligned}
		\chi_1(t,\boldsymbol{n})  \stackrel{\text{long times}}{\sim} &\frac{e^{-\Gamma t/2}}{\eta^2}\Bigg\lvert \frac{1}{64} \left(64g^2\sinh^2(\eta t) n_2 + 4\left(-(\Gamma_1-\Gamma_2)\sinh^2(\eta t)  + 4\eta \cosh(\eta t)\right)^2n_1\right)\\
		&-\int \frac{d\epsilon}{2\pi}\Bigg[\frac{e^{-\Gamma t_0/4}e^{-\eta(t-t_0)}\left[(\Gamma_1-\Gamma_2+4\eta)^2f_1\Gamma_1+16g^2f_2\Gamma_2\right]\cos\left((\epsilon-\epsilon_d)(t-t_0)\right)}{2\left(16(\epsilon-\epsilon_d)^2 +(\Gamma+4\eta)^2 \right)}\\
		&+\frac{e^{-\Gamma t_0/4}e^{\eta(t-t_0)}  \left[(\Gamma_2-\Gamma_1+4\eta)^2f_1\Gamma_1+16g^2f_2\Gamma_2\right] \cos\left((\epsilon-\epsilon_d)(t-t_0)\right)}{2\left(16(\epsilon-\epsilon_d)^2 +(\Gamma-4\eta)^2 \right)}\Bigg]
		\Bigg\rvert;
	\end{aligned}
\end{equation}
\begin{equation}
	\begin{aligned}
		\chi_1^{\text{\tiny{EP}}}(t,\boldsymbol{n}^{\text{\tiny{EP}}})\stackrel{\text{long times}}{\sim}&e^{-\Gamma t/2}\Bigg\lvert\left(1+2t g_{\text{\tiny{EP}}}\right)n_1^{\text{\tiny{EP}}} + t^2 g^2_{\text{\tiny{EP}}}(n_1^{\text{\tiny{EP}}} + n_2^{\text{\tiny{EP}}})\\
		&+e^{\Gamma t_0/4}\int \frac{d\epsilon}{2\pi} 4 \Bigg[-\frac{1}{2} \frac{\cos\left[\left(\epsilon-\epsilon_d \right)\left(t-t_0\right) \right]}{\left(\frac{\Gamma}{4}\right)^2 + (\epsilon-\epsilon_d)^2} -g_{\text{\tiny{EP}}}\frac{t\cos\left[\left(\epsilon-\epsilon_d \right)\left(t-t_0\right)\right]}{\left(\frac{\Gamma}{4}\right)^2 + (\epsilon-\epsilon_d)^2}\Bigg]\Bigg\rvert.
	\end{aligned}
\end{equation}In the ratio $\chi_1^{\text{\tiny{EP}}}(t,\boldsymbol{n}^{\text{\tiny{EP}}})/\chi_1(t,\boldsymbol{n})$, the time dependence through $e^{-\Gamma t/2}$ cancels. Moreover, at long times with $\eta>0$ (``overdamping"), $e^{\eta t}$ dominates over $e^{-\eta t}$. Similar results can be obtained for $\chi_2$. Therefore, the ratio shows the following time scaling,
\begin{equation}
	\begin{aligned}
		\frac{\chi_j^{\text{\tiny{EP}}}(t)}{\chi_j(t)}\sim \frac{\mathcal O(t^2)}{\mathcal O(e^{\eta t})} <1
	\end{aligned}
\end{equation}Therefore, at long times, the system operating at an EP is closer to its steady state than the system at non-EPs with $\eta>0$. Therefore, the EP corresponds to the point of critical damping, as seen in a classical damped harmonic oscillator - it is the point separating oscillatory and non-oscillatory dynamical regimes, and represents the fastest non-oscillatory approach to the steady state.

\section{Beyond the double quantum dot}
\label{app:C}

We now consider the case of an $N$-dot chain, with each dot connected to a fermionic thermal reservoir. For $N>2$, this model in general features higher-order EPs in both Heisenberg and master-equation approaches. The interpretation of such EPs, specifically with respect to the dynamics of the system, is a challenging task \cite{Khandelwal2021}. Moreover, for larger $N$, the operators may have unfactorable characteristic polynomials of degree greater than 4, which may mean that analytical closed form expression of eigenvalues cannot be determined. 

Let us first consider the Heisenberg approach. For our model with $N$ dots, each connected with its own thermal reservoir, the matrix $A$ is a $N\times N$ tridiagonal matrix with uniform off-diagonal entries and non-uniform diagonal entries,
\begin{align}
	A_N = -\begin{pmatrix}
		\frac{\Gamma_1}{2}+i\epsilon_d &ig &0 &&\dots &0\\
		ig & \frac{\Gamma_2}{2}+i\epsilon_d &ig &&\dots &0 \\
		\vdots &&\ddots  \\
		0 &\dots &&&ig & \frac{\Gamma_N}{2}+i\epsilon_d
	\end{pmatrix}
\end{align}For completely non-uniform diagonal entries, there is no known analytical closed-form expression for the eigenvalues of the above matrix. Moreover, the same holds for boundary-driven systems (i.e., a chain of quantum dots with reservoirs attached only at the ends). It can further be checked that if all couplings to reservoirs are equal, $A$ is a uniform tridiagonal as well as Toeplitz matrix, and cannot show EPs due to the form of its eigenvalues \cite{Noschese2013}. Therefore, we consider the minimal complication to this model such that $\Gamma_j = \Gamma_1$ for odd $j$ and $\Gamma_j = \Gamma_2$ for even $j$, giving us a two-periodic diagonal. If $N=2d$ for some $d\in\mathds N$ (i.e., $N$ is even), the eigenvalues of such a matrix are given by \cite{Dyachenko2022},
\begin{align}
	\sigma(A_{2d}) = \left\{-i\epsilon_d -\frac{\Gamma}{4} \pm \eta_{N}^{(j)} \right\}_j, \quad j=1,2,..,d
\end{align}where $\eta_{N}^{(j)}=\sqrt{\lambda_j^2 + \left(\frac{\Gamma_1-\Gamma_2}{4} \right)^2}$. 
$\lambda_j$ are the $N$ eigenvalues of $A_N$ when the diagonal entries are zero and are given by $\lambda_j =-2ig\cos\left(\frac{j\pi}{N+1} \right)$ \cite{Noschese2013}. If $N=2d+1$ (i.e., $N$ is odd), the spectrum is given by $\sigma(A_{2d+1}) = \left\{-i\epsilon_d -\frac{\Gamma}{4} \pm \eta_{N}^{(j)} \right\}_j \cup\left\{-\Gamma_1/2-i\epsilon_d\right\}$. It can be verified that there are second-order EPs for all $\eta_{N}^{(j)}=0$ ($j=1,2,\cdot\cdot\cdot,d$). In particular, for $N=2$, the model reduces to the one considered in the main text, i.e., $\eta_2^{(j)}=\eta$, and we obtain a single second-order EP. For $N=3$, we obtain
\begin{align}
	\sigma(A_{3}) = \left\{-\Gamma_1/2-i\epsilon_d,-i\epsilon_d -\frac{\Gamma}{4} \pm \eta_{3} \right\},
 \end{align}with $\eta_3\equiv\eta_3^{(j)} = \sqrt{ \left(\frac{\Gamma_1-\Gamma_2}{4} \right)^2-2 g^2 }$. In the case $\Gamma_2=0$, we obtain a boundary-driven three-dot chain. It can be verified that in this case, the above eigenvalues coincide with those in Eq. (13) in the main text.

Now, let us compare the above with the ME approach. Closed-form expressions of eigenvalues for the general scenario are an open problem \cite{Prosen2008,Prosen2010}. We therefore consider the minimal, three-dot case, along with the above simplification of alternating couplings. Considering a local master equation as in the main text,
\begin{align}
	\dot\rho = \mathcal L_3\rho = -i[\hat H,\rho] &+ \Gamma_1^+ f_1(\epsilon_d)\mathcal D[\hat \sigma_\alpha^{(1)}]\rho+  +\Gamma_2^+ f_2(\epsilon_d)\mathcal D[\hat \sigma_+^{(2)}]\rho +\Gamma_1^+\mathcal D[\hat \sigma_+^{(3)}]\rho\\
	&+\Gamma_1^-(1- f_1(\epsilon_d))\mathcal D[\hat \sigma_\alpha^{(1)}]\rho +\Gamma_2^- (1-f_2(\epsilon_d))\mathcal D[\hat \sigma_-^{(2)}]\rho +\Gamma_1^- (1-f_3(\epsilon_d))\mathcal D[\hat \sigma_-^{(3)}]\rho,
\end{align}with $\hat H = \epsilon_d\sum_j\sigma_+^{(j)}\sigma_-^{(j)}+g(\sigma_+^{(1)}\sigma_-^{(2)}+\sigma_-^{(1)}\sigma_+^{(2)}) + g(\sigma_+^{(2)}\sigma_-^{(3)}+\sigma_-^{(2)}\sigma_+^{(3)})$. Four relevant eigenvalues of the above Liouvillian are given by
\begin{align}
	\sigma(\mathcal L_3) = \left\{-\frac{3\Gamma}{4}-\frac{\Gamma_1}{2}\pm\eta_3,-\frac{\Gamma}{4} -\frac{\Gamma_1}{2}\pm\eta_3 \right\},
\end{align}with $\Gamma=\Gamma_1+\Gamma_2$. The Liouvillian shows two second-order EPs at $\eta_3=0$. The parameter $\eta_3$ is identical in both Heisenberg  and ME approaches. Therefore, as in the case of the DQD, we find that the EPs in both approaches are equivalent.

\end{document}